\documentclass{ptapap}

\usepackage{graphicx}

\author{Alexandre David-Uraz}[UD,FIT]
\author{Gregg Wade}[RMC]
\author{Anthony Moffat}[UdeM]
\author{Stan Owocki}[UD]
\author{V\'{e}ronique Petit}[UD]
\author{the BRITE team}
\affil[UD]{Bartol Research Institute, University of Delaware\\
  Newark, DE 19716, USA}
\affil[FIT]{Department of Physics \& Space Sciences, Florida Institute of Technology\\
  Melbourne, FL 32901, USA}
\affil[RMC]{Department of Physics \& Space Science, Royal Military College of Canada\\
  PO Box 17000, Stn Forces, Kingston, ON K7K 4B4, Canada}
\affil[UdeM]{Département de physique and Centre de Recherche en Astrophysique du Québec, Université de Montréal\\
  C.P. 6128, Succursale Centre-Ville Montréal, QC H3C 3J7 Canada}
  
\renewcommand{\thefootnote}{\fnsymbol{footnote}}

\title{Rotationally modulated photometric variations in B supergiants?\footnotemark[1]}

\begin{document}

\maketitle

\footnotetext[1]{Based on data collected by the BRITE Constellation satellite mission, designed, built, launched, operated and supported by the Austrian Research Promotion Agency (FFG), the University of Vienna, the Technical University of Graz, the Canadian Space Agency (CSA), the University of Toronto Institute for Aerospace Studies (UTIAS), the Foundation for Polish Science \& Technology (FNiTP MNiSW), and National Science Centre (NCN).}

\renewcommand{\thefootnote}{\arabic{footnote}}

\begin{abstract}

In this contribution, we present BRITE observations of the early-B supergiants $\epsilon$ Ori and $\kappa$ Ori. We perform a preliminary analysis of the data acquired over the first two Orion observing runs. We evaluate whether they are compatible with co-rotating bright spots and discuss the challenges of such an approach.

\end{abstract}

\section{Introduction}

The advent of high-precision space-based broadband optical photometry with satellites (or ensembles of satellites) such as the \textit{Microvariability and Oscillations of STars} (MOST; \citealt{2003PASP..115.1023W}) and \textit{BRIght Target Explorer} (BRITE; \citealt{2014PASP..126..573W}) missions has opened the door to a brand new picture of the photospheres of hot massive stars. In particular, recent studies have led to the detection of co-rotating bright spots on the surface of a few O stars \citep{2014MNRAS.441..910R, 2017arXiv171008414R}. The existence of such spots has been proposed \citep{1996ApJ...462..469C, 2017MNRAS.470.3672D} to explain the formation of \textit{co-rotating interaction regions} (CIRs), which in turn are postulated \citep{1986A&A...165..157M} to lead to recurring \textit{discrete absorption components} (DACs) which migrate through the velocity space of the absorption troughs of ultraviolet resonance lines, as revealed in timeseries of spectra obtained by the \textit{International Ultraviolet Explorer} (IUE; e.g., \citealt{1989ApJS...69..527H, 1996A&AS..116..257K}). While the physical origin of these bright spots remains contested, one popular hypothesis contends that they are caused by small-scale magnetic fields which can be generated in the subsurface convection zone due to the iron opacity bump (FeCZ; \citealt{2011A&A...534A.140C}).

$\epsilon$ Ori and $\kappa$ Ori are two bright early B supergiants (respectively B0Ia and B0.5Ia) which have been observed by BRITE. Together, they constitute an interesting testbed for the study of photospheric perturbations such as bright spots, for a few reasons. First, given their magnitudes (respectively, $m_V$ = 1.69 and $m_V$ = 2.06; \citealt{2002yCat.2237....0D}), they are prime candidates for high signal-to-noise ratio (SNR) observations, allowing us to place very tight constraints on their properties. Secondly, their current evolutionary stage is of interest for the study of spots\footnote{It should be noted, however, that B stars are also known to be the theatre of various forms of variability, from SPB/$\beta$ Cephei pulsations to rotational modulation \citep{2011MNRAS.413.2403B}.} as the envelopes of hot supergiants are expected to host more convection than their main sequence counterparts \citep{2009A&A...499..279C}. Finally, if one naively posits that the properties of putative bright spots should be intimately related to stellar parameters, it would therefore be expected that if $\epsilon$ Ori and $\kappa$ Ori show signatures of co-rotating bright surface spots, these spots would have similar characteristics, an assertion that we can test. Conversely, any departure from that expectation informs us about the nature of these photospheric structures.

$\epsilon$ Ori is a known variable star. \citet{2002A&A...388..587P} have traced the evolution of a DAC in its ultraviolet lines for at least 17h, and periods ranging from about 0.8 to 19 days have been recovered from its optical spectra (e.g., \citealt{2013AJ....145...95T}). From these, a rotational period of either $\sim$4 days or $\sim$18 days is inferred. From its radius ($24.0 R_\odot$; \citealt{2006A&A...446..279C}) and its projected rotational velocity (60 km/s; \citealt{2014A&A...562A.135S}), we derive a maximum rotational period of about 20 days. On the other hand, repeating the same calculation for $\kappa$ Ori ($ R_* = 22.2 R_\odot$; $v \sin i = 54$ km/s), we obtain a maximum period of about 21 days.

\section{Observations}

So far, the Orion field has been observed five times by BRITE (2013, 2014, 2015, 2016 and 2017) in both red and blue wavebands. However, we focus here on the first two observing runs. The details of the observations are presented in Table~\ref{tab:obsrun}.

\begin{table}
\caption{Details of the two Orion observing runs presented in this study. The satellites are UniBRITE (UBr; red waveband), BRITE Austria (BAb; blue waveband), BRITE Lem (BLb; blue waveband), BRITE Heweliusz (BHr; red waveband) and BRITE Toronto (BTr; red waveband).}\label{tab:obsrun}
\begin{tabular}{lllll}
Run & Starting date & Total length (days) & Telescopes\\
Orion I & 2013-11-07 & 131 & UBr, BAb\\
Orion II & 2014-09-24 & 174 & BAb, BLb, BTr, BHr\\
\end{tabular}
\end{table}

\begin{figure}
\includegraphics[width=\textwidth]{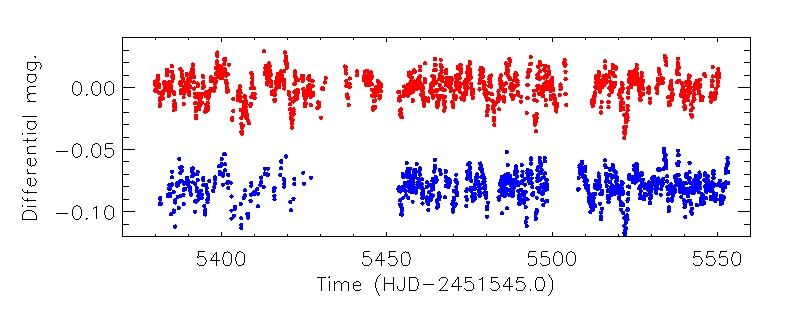}
\caption{Orion II light curves (red and blue filters) of $\epsilon$ Ori. The blue light curve is shifted downwards. We can see that both light curves exhibit similar variations, with a maximum amplitude of roughly 30 mmag. However, no obvious repeatable pattern is observed.}
\label{fig:lc}
\end{figure}

Sample light curves are shown in Fig.~\ref{fig:lc}. The main characteristic that we can observe (for both stars) is that there appear to be significant variations with a maximum amplitude of roughly 30 mmag. These variations do not exhibit an obvious periodic behaviour. The point-to-point precision is of millimagnitude order.

\section{Preliminary analysis}

\begin{figure}
  \centering
  \begin{minipage}{0.48\textwidth}
    \includegraphics[width=\textwidth]{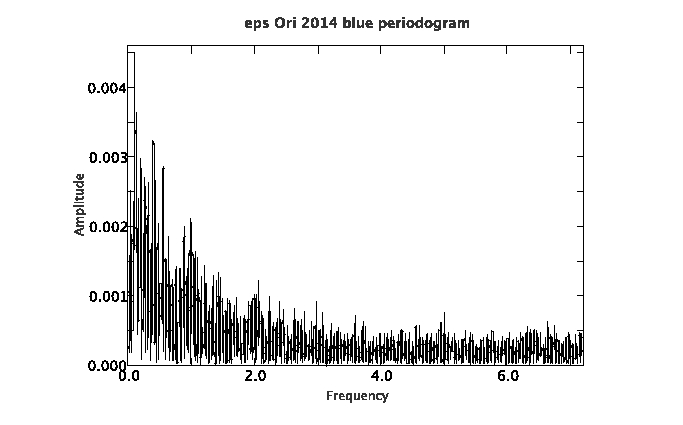}
    \caption{Periodogram of the blue light curve of $\epsilon$ Ori from the Orion II run; we can see groupings of frequencies around 0.25 c/d, 0.5 c/d and 1 c/d.}
    \label{fig:ft}
  \end{minipage}
  \quad
  \begin{minipage}{0.48\textwidth}
    \includegraphics[width=\textwidth]{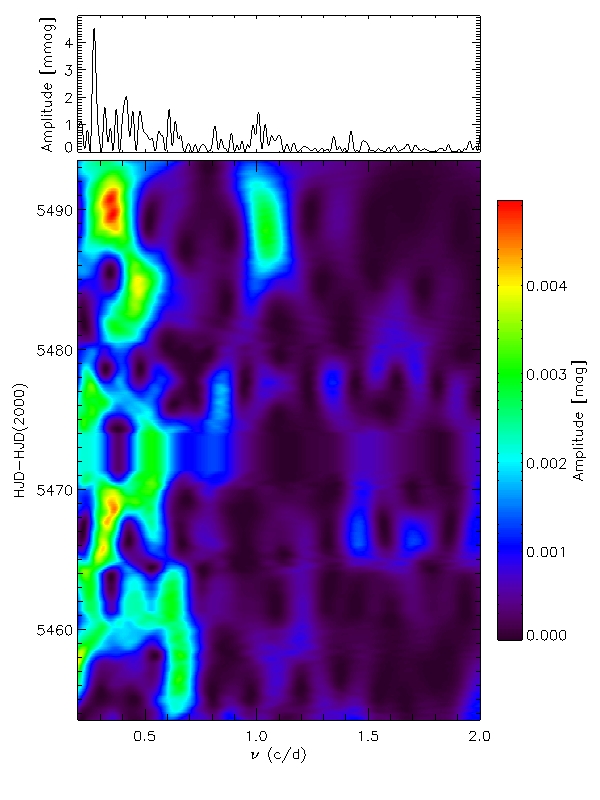}
    \caption{Time-frequency analysis of a portion of about 30 days of the Orion II blue light curve of $\epsilon$ Ori performed with an 8-day window; frequencies can notably be found around 0.25 c/d, 0.5 c/d and 1 c/d. The integrated periodogram is shown on the top.}
    \label{fig:stft}
  \end{minipage}
\end{figure}

We perform a period search on these light curves to find any periodicity. While neither star shows clear, periodic variations, various frequencies are detected (see Fig.~\ref{fig:ft}), and a time-frequency analysis suggests that these frequencies appear and disappear over time. In the case of $\epsilon$ Ori, clumps of frequencies around $\sim$0.25 c/d, $\sim$ 0.5 c/d, $\sim$0.75 c/d and $\sim$1 c/d (therefore, roughly the first few integer multiples of a base frequency of around 0.25 c/d) are detected at any given time. At face value, this seems somewhat consistent with the type of observational signatures historically associated with the bright spot/co-rotating interaction regions (CIR) phenomenology (e.g., \citealt{2011ApJ...735...34C}), in which case this base frequency could be of rotational origin (meaning that $P \simeq 4$ days, as previously suggested as one of the possible periods by optical spectra). The result of this analysis on a portion of the blue Orion II light curve is shown in Fig.~\ref{fig:stft}.

However, more analysis is required in order to lend credence to the bright spot scenario. In particular, pulsations must first be investigated in depth before ruling them out as being responsible for the observed variability. A similar analysis was performed on the light curves of $\kappa$ Ori, revealing frequencies at around 0.4 c/d and 1.2 c/d. While it is too early to conclude whether these periods are due to rotational modulation, it should be noted that if the base frequencies of $\epsilon$ Ori and $\kappa$ Ori are indeed of rotational origin, both stars can then be inferred to be viewed at a rather small inclination, which might be problematic.

\section{Conclusions and future work}

While this study remains in the early stages and much more work is required, preliminary results show pleasant parallels with the observational signatures typically ascribed to rotational modulation due to co-rotating bright surface spots. If this scenario is favored, these observations can help us learn a great deal about the properties and the nature of such photospheric perturbations. In particular, the bright spot/magnetic spot connection can hopefully be further investigated, as both stars are ideal candidates for ultra-deep magnetometry given their magnitude and low projected rotational velocity (e.g., \citealt{2016MNRAS.456....2W}). Furthermore, advances in modelling will also prove invaluable in constraining the properties of bright spots and associated wind structures; building on our recent work \citep{2017MNRAS.470.3672D}, the next logical step will be to produce hydrodynamical models of co-rotating interaction regions in three dimensions, studying, among other things, the effects of inclination on observational signatures. Meanwhile however, a more in-depth period analysis, including the data from all 5 BRITE Orion runs, will be necessary to more robustly establish whether the observed variations are indeed consistent with bright spots.

\acknowledgements{ADU gratefully acknowledges support from the \textit{Fonds qu\'{e}b\'{e}cois de la recherche en nature et technologies}.}

\bibliographystyle{ptapap}
\bibliography{ptapapdoc_adu}

\end{document}